\documentclass[11pt]{article}
\usepackage[top = 2 cm, bottom = 2.3 cm, left = 2.5 cm, right = 1.8 cm]{geometry}
\usepackage{authblk, cite, color, amssymb, amsmath, enumitem}
\usepackage[hypertex, colorlinks = true, linkcolor = blue, citecolor = red]{hyperref}
\usepackage[dvipsnames]{xcolor}

\begin{document}

\title{The width of a black hole atmosphere}

\author[1,2]{M. Gogberashvili \thanks{gogber@gmail.com}}
\author[1]{T. Tsiskaridze \thanks{tinatin.tsiskaridze@gmail.com}}
\affil[1]{Javakhishvili State University, 3 Chavchavadze Ave., Tbilisi 0179, Georgia}
\affil[2]{Andronikashvili Institute of Physics, 6 Tamarashvili St., Tbilisi 0177, Georgia}

\maketitle

\begin{abstract}

Black holes, as classical solutions of General Relativity, are expected to exhibit quantum properties near their horizons. In this paper, we examine the behavior of quantum particles near the Schwarzschild horizon by solving the Klein-Gordon equation in the quasi-classical approximation. Our analysis shows that, rather than the periodic-in-time solutions typically associated with black hole exteriors or interiors, particles near the horizon exhibit exponentially decaying (or growing) time-dependent solutions with a complex phase. This suggests that particles are unlikely to cross the Schwarzschild horizon; instead, they are either absorbed by the black hole or reflected back, forming a thin atmospheric layer around the horizon. Using geodesic equations derived from the Klein-Gordon equation, we estimate the width of this atmosphere.

\vskip 3mm
\noindent
PACS numbers: 04.70.-s (Physics of black holes); 04.20.Dw (Singularities and cosmic censorship); 04.62.+v (Quantum fields in curved spacetime)

\vskip 3mm
\noindent
Keywords: Black hole atmosphere, Schwarzschild horizon, Klein-Gordon equation in quasi-classical approximation
\end{abstract}


\section{Introduction}

Black holes (BHs) are classical solutions of General Relativity \cite{BH-1, BH-2, BH-3, Wald, gravMTW}, but they can also be understood as macroscopic quantum objects. Inside a BH, quantum gravitational effects are expected to dominate, potentially enabling a regular description of spacetime within a quantum framework. Observationally, the interior of a BH remains inaccessible, but the possibility of BHs without singularities has been suggested \cite{Ansoldi:2008jw, Cardoso:2019rvt}.

As BHs result from gravitational collapse, compressing matter to a minimum size without violating quantum theory \cite{Malafarina:2017csn, Kerr:2023rpn}, they can be considered horizon-sized massive spheres. This requires revising the classical depiction of BHs, not just at the central singularity but also at the horizon scale \cite{Modif-1, Modif-3}. The perspective that the interior of a BH contains a physical object rather than empty space with a central singularity helps reconcile two competing viewpoints: one suggesting that the energy density at the horizon is small \cite{Howard:1984qp}, and the other proposing the existence of structures such as a "brick wall" \cite{tHooft:1984kcu}, "fuzzball" \cite{Mathur:2005zp}, or "firewall" \cite{Modif-2}, which induce significant back-reactions on the geometry.

If BHs are indeed `normal' astrophysical objects, their event horizons would enclose an impenetrable sphere of ultra-dense `quantum' matter, potentially leading to observable phenomena such as wave reflections from the event horizon \cite{meG1, meG2, meG3, Kuchiev:2003fy}. The reflected waves would form an 'atmosphere,' the thickness of which has been estimated using the concept of BH entropy \cite{Ghosh:2009xg}. In this paper, we aim to investigate the quantum behavior of matter near the Schwarzschild horizon, focusing on the formation of a BH atmosphere and estimating its width using the Klein-Gordon equation in the quasi-classical approximation.

The paper is organized as follows. In Sec.~\ref{Sch-metric}, we discuss the challenges in realizing Schwarzschild BHs. In Sec.~\ref{KG-Sch}, we solve the Klein-Gordon equation near the Schwarzschild horizon. The solutions obtained include real-valued, exponentially time-dependent factors with complex phases, suggesting that particles likely do not traverse the BH's event horizon but are instead absorbed or partially reflected. In Sec.~\ref{BH atmosphere}, we estimate the width of the BH atmosphere formed by reflected particles using geodesic equations derived from the Klein-Gordon equation in the quasiclassical approximation. Finally, Sec.~\ref{Concl} presents concluding remarks and discussions.


\section{Schwarzschild singularities} \label{Sch-metric}

The most important solution in General Relativity is the Schwarzschild metric:
\begin{equation} \label{Schwarzschild}
ds^2 = f(r)\, dt^2 - \frac {dr^2}{f(r)} - r^2 d\theta^2 - r^2 \sin^2 \theta d\phi^2~,
\end{equation}
where the metric function
\begin{equation} \label{f(r)}
f(r) = 1 - \frac {r_s}{r}
\end{equation}
contains the parameter $r_s = 2GM$, which determines the Schwarzschild horizon. Here, $G$ denotes the gravitational constant, and $M$ is the mass of the spherical object.

For Schwarzschild BHs, the geometry described by \eqref{Schwarzschild} is typically assumed to remain valid even inside the horizon ($r < r_s$). In this case, the mass $M$ is not localized at a specific point (since the stress-energy-momentum tensor is zero) but serves as a useful parameter for describing a BH that has not yet fully formed from the perspective of a distant observer. After complete collapse, all matter that formed the BH is expected to vanish into the central singularity at $r = 0$, leaving behind only the vacuum gravitational field \cite{BH-1, BH-2, BH-3, Wald, gravMTW}.

In addition to the central singularity, the Schwarzschild solution \eqref{Schwarzschild} also features a horizon peculiarity at $r = r_s$, leading to divisions by zero or multiplications by infinity in certain geometrical quantities. However, the point $r = r_s$ is regarded as a coordinate singularity that can be eliminated by switching to appropriate coordinates. In contrast, the point $r = 0$ corresponds to a true physical singularity, as evidenced by coordinate-independent quantities such as the Kretschmann invariant. For sufficiently large BHs, the Kretschmann invariant can be made arbitrarily small as $r \to r_s$. However, the claim that some characterizing quantities, such as the metric determinant, Ricci scalar, or Kretschmann invariant, remain finite at $r = r_s$ relies on the mathematically questionable assumption of mutual cancellations of delta-function-like divergences \cite{Colombeau-1, Colombeau-2, Gel-Sch, Heinzle:2001bk, Foukzon, Gogberashvili:2024ogp}.

To address the issues posed by singularities, distributional approaches for metrics can be utilized \cite{Colombeau-1, Colombeau-2, Gel-Sch, Heinzle:2001bk, Foukzon, Gogberashvili:2024ogp}. Consider the regularized Schwarzschild radial variable near the horizon:
\begin{equation} \label{r+epsilon}
r \quad \to \quad r_\epsilon = r_s + \sqrt{\left( r - r_s \right)^2 + \epsilon^2}~,
\end{equation}
where $\epsilon$ is an infinitesimal length parameter. In this coordinates, the Ricci scalar and Kretschmann invariant exhibit singular behavior at the horizon:
\begin{equation}
\begin{split}
{\mathcal R}_\epsilon \big|_{\epsilon \to 0} \approx \frac {\epsilon^2}{r_s\left[\left( r - r_s \right)^2 + \epsilon^2\right]^{3/2}}\bigg|_{\epsilon \to 0} \sim \delta (r - r_s) \quad \to \quad \infty~, \\
{\mathcal R}^{\alpha\beta\gamma\delta}{\mathcal R}_{\alpha\beta\gamma\delta} \big|_{\epsilon \to 0} \approx \frac {12 r_s^2}{r^6} + \frac {4\epsilon^4}{r_s^4\left[\left( r - r_s \right)^2 + \epsilon^2\right]^3}\bigg|_{\epsilon \to 0} \quad \to \quad \infty~.
\end{split}
\end{equation}
Thus, the apparent smallness of these functions at the BH horizon does not imply that the curvature is small \cite{Ha}. This is because three out of the six nonzero independent components of the mixed Riemann tensor for the Schwarzschild metric \eqref{Schwarzschild} diverge as $r \to r_s$.

The conventional approach, which treats $r = r_s$ in \eqref{Schwarzschild} as merely a coordinate singularity, is the primary motivation for using alternative radial variables instead of $r$. This ensures that the horizon singularity is `eliminated', allowing classical particles to traverse the horizon and reach the central naked singularity without obstruction. A key element of all singular coordinate transformations of the Schwarzschild metric is the Regge-Wheeler tortoise coordinate \cite{BH-1, BH-2, BH-3, Wald, gravMTW}:
\begin{equation} \label{tortoise}
r^* = \int  \left( 1 - \frac {r_s}{r}\right)^{-1}dr = r + r_s \ln \left( \frac {r}{r_s} - 1\right)~. \qquad \left \{
\begin{array} {lr}
&r_s < r < \infty \\
&- \infty < r^* < \infty
\end{array}
\right.
\end{equation}
If the domain of definition for the factor $1/(r-r_s)$ includes the singular point $r = r_s$, as is often assumed for the tortoise coordinate in \eqref{tortoise}, it becomes necessary to introduce generalized derivatives.
\begin{equation} \label{tortoise-gen'}
\frac {d r^*}{dr} = \frac {r}{r - r_s} + \mathcal{B} \, \delta (r-r_s)~,
\end{equation}
where $\mathcal{B}$ is a constant. Then, the solution of \eqref{tortoise-gen'}, representing the generalized Regge-Wheeler tortoise coordinate \eqref{tortoise}, is given by:
\begin{equation} \label{tortoise'}
r^* =  r + r_s \ln \left( \frac {r}{r_s} - 1\right) + \mathcal{B} \, H(r - r_s)~,
\end{equation}
incorporates the Heaviside function (see, for instance, (3.57) in \cite{Generalize}).

In the class of generalized functions \eqref{tortoise'}, singular coordinate transformations of \eqref{Schwarzschild} (such as those considered by Kruskal-Szekeres, Eddington-Finkelstein, Lema\^{\i}tre, or Gullstrand-Painlev\'{e}) yield $\delta$-functions appearing in the second derivatives, as these transformations involve factors like $\sqrt{r_s - r}$ or $\ln |r_s - r|$ \cite{meG4}. As a result, the transformed metric tensors are not differentiable at $r = r_s$ and the vacuum Einstein equation for these metrics is modified by $\delta$-sources on Schwarzschild sphere.

Moreover, the classical action for particles in a Schwarzschild background exhibits a logarithmic singularity at the horizon \cite{Kuchiev:2003fy}, which is a covariant property that persists even in coordinates that eliminate the singularity of the metric and the classical equations of motion. Therefore, in equations for matter fields, the point $r = r_s$ should be excluded from the definition interval for the Regge-Wheeler tortoise coordinate \eqref{tortoise}, and appropriate boundary conditions must be imposed at the horizon.


\section{Klein-Gordon equation near the horizon} \label{KG-Sch}

Since Einstein's gravity is insensitive to particle spin, the study of strong gravitational fields near the Schwarzschild horizon can be simplified by focusing on the Klein-Gordon equation
\begin{equation} \label{wave}
(\Box + \mu^2)\Phi = \left[\frac {1}{ \sqrt{-g}}\partial_\mu \left( \sqrt{-g}g^{\mu\nu}\partial_\nu \right) + \mu^2 \right]\Phi = 0~.
\end{equation}
For photons, we have $\mu = 0$, whereas for fermions, a corresponding equation of this form must be derived from the first-order Dirac system. The Schwarzschild metric \eqref{Schwarzschild} is highly symmetric, allowing for variable separation:
\begin{equation} \label{Phi}
\Phi \sim \psi (t, r) Y_{lm}(\theta, \phi) ~,
\end{equation}
where $Y_{lm}(\theta, \phi)$ are spherical harmonics. Substituting this into the Klein-Gordon equation \eqref{wave} yields:
\begin{equation} \label{psi}
\bigg \{ r^2 \partial_t^2 - f \partial_r\left( r^2 f\partial_r\right) + f\big[l(l+1) + r^2 \mu^2 \big]\bigg \}\psi (t,r) = 0~,
\end{equation}
where $f(r)$ is the metric function defined in \eqref{f(r)} and $l$ is the orbital quantum number. Near the horizon, where $f \to 0$, the terms in \eqref{psi} containing mass and angular momentum become negligible:
\begin{equation} \label{m,l=0}
\mu, l \to 0~.
\end{equation}
Further separation of variables,
\begin{equation} \label{psi=RT}
\psi (t,r) = \frac 1r T(t)R(r) ~,
\end{equation}
reduces \eqref{psi} to the system:
\begin{equation} \label{T''}
\partial^2_t T = \mathcal{C} T ~,
\end{equation}
\begin{equation} \label{R''}
f^2 \partial_r^2 R + \frac{r_sf}{r^2} \partial_r R - \left(\mathcal{C} + \frac {r_sf}{r^3}\right) R = 0~,
\end{equation}
where $\mathcal{C}$ is a separation constant.

To analyze the behavior of quantum particles near the horizon, one must solve the system \eqref{T''}–\eqref{R''} with appropriate boundary conditions at $f = 0$. In the standard approach, the boundary conditions for \eqref{T''}–\eqref{R''} are typically imposed by assuming the presence of horizon-crossing radial waves \cite{Matz, Star}:
\begin{equation} \label{solution-f=0}
T(t)_{f \to 0} \sim  e^{\pm i\omega t}~, \qquad R(r)_{f \to 0} \sim e^{\pm i\omega r^*} ~,
\end{equation}
where $r^*$ is the Regge-Wheeler tortoise coordinate \eqref{tortoise}. This transformation reduces \eqref{R''} to a Schr\"{o}dinger-type  equation,
\begin{equation} \label{KG-R}
\frac {d^2R}{dr^{*2}} - \left[V(r^*) + \mathcal{C}\right] R = 0 ~,
\end{equation}
where in the simplest case \eqref{m,l=0} the effective potential has the form:
\begin{equation} \label{Veff}
V(r^*) = \frac {r_sf(r^*)}{r^3(r^*)} ~.
\end{equation}
The conjecture \eqref{solution-f=0} correlates with a negative separation constant in the equations \eqref{T''}–\eqref{R''},
\begin{equation}
\mathcal{C} = -\omega^2 < 0 ~.
\end{equation}
However, as previously noted, the transformation \eqref{tortoise} is singular, and the solution \eqref{solution-f=0} does not satisfy \eqref{R''} due to the appearance of a Dirac delta function in the second derivatives of the tortoise coordinate function \eqref{tortoise'} at the horizon, $r = r_s$.

To determine the correct boundary conditions at the BH horizon, we use $f(r)$ as an independent variable in the region $r_s \le r \le \infty$ ($0 \le f \le 1$) and rewrite \eqref{R''} in the form:
\begin{equation} \label{R(f)}
f^2(1-f)^4 R'' + f(1-f)^3(1-3f) R' -\left[r_s^2\mathcal{C} + f(1-f)^3 \right] R = 0~,
\end{equation}
where primes denote derivatives with respect to $f$. Near the horizon, this equation simplifies to the asymptotic form:
\begin{equation} \label{R(f=0)}
f^2R'' + f R' - r_s^2\mathcal{C} R = 0~. \qquad (f \to 0)
\end{equation}
It is evident that as $f \to 0$, the first two terms in \eqref{R(f=0)} vanish, leaving only the third term. Thus, for a regular solution, the boundary condition must satisfy:
\begin{equation} \label{R=0}
R|_{f \to 0 } \to 0~,
\end{equation}
implying that the probability of particles crossing the horizon approaches zero. Consequently, infalling particles are either absorbed by the BH or reflected off its horizon \cite{Gogberashvili:2016xcx}.

Now, let us seek a solution to \eqref{R(f)} that satisfies the appropriate boundary condition \eqref{R=0} using the Frobenius method:
\begin{equation} \label{R=}
R (f) = \sum_{i}^{\infty} a_{i} f^{i}~, \qquad (i = 1, ...,\infty)
\end{equation}
where $a_{i}$ are constants.  Substituting this expansion into \eqref{R(f)} leads to an algebraic system for the coefficients of $f^i$, which must vanish term by term. The first equation of this system ($i = 1$),
\begin{equation} \label{a}
f(1-f)^3(1-3f)a_1 -\left[r_s^2\mathcal{C} + f(1-f)^3 \right]a_1f \approx \left(1 - r_s^2 \mathcal{C} \right)\left(a_1f\right) = 0 ~,
\end{equation}
implies that the separation constant in the system \eqref{T''}–\eqref{R''} is positive \cite{Gogberashvili:2016xcx, Qin}:
\begin{equation} \label{k}
\mathcal{C} \approx \frac {1}{r_s^2} > 0 ~.
\end{equation}
From \eqref{T''}, we then obtain a real-valued exponentially time-dependent solution, with a complex phase:
\begin{equation} \label{T}
T(t)= T_0 e^{\pm t/r_s} ~.
\end{equation}
Thus, in the Schwarzschild coordinates of a distant observer, the wave function \eqref{psi=RT} takes the form \cite{Gogberashvili:2016xcx}:
\begin{equation} \label{psi=}
\psi(t,r) \sim \frac {e^{\pm t/r_s}}{r} \sum_{i}^{\infty} a_{i} \left(1 - \frac {r_s}{r}\right)^{i} ~.
\end{equation}
The constants $a_i$ satisfy the condition:
\begin{equation}
\frac{a_{i+1}}{a_i}\big|_{i \to \infty} \to 2~,
\end{equation}
which ensures convergence of the radial wave function \eqref{R=}:
\begin{equation}
\frac{a_{i+1} f^{i+1}}{a_i f^i}\big|_{i \to \infty} < 1 ~,
\end{equation}
leading to the constraint:
\begin{equation} \label{f<1/2}
0 < f < \frac 12~. \qquad (r_s <r < 2r_s)
\end{equation}
Thus, the solution \eqref{psi=} is applicable in the region of interest close to the Schwarzschild horizon.


\section{Black hole atmosphere} \label{BH atmosphere}

The obtained solution with the complex phase \eqref{psi=} of the matter field equations near the Schwarzschild horizon differs significantly from the familiar internal \cite{BH-wave-1, BH-wave-2} and outer \cite{BH-out} periodic-in-time solutions \eqref{solution-f=0} for particles in a BH field. In semiclassical quantum mechanics, a wave function with a complex phase typically represents a tunneling process through a potential barrier \cite{Tunneling-1, Tunneling-2}, suggesting that particle penetration through the BH horizon is classically forbidden. Thus, our exponentially decaying (or growing) time-dependent solutions \eqref{T} imply that quantum particles are likely absorbed or reflected by the Schwarzschild horizon, rather than crossing it \cite{Gogberashvili:2017xti, Beradze:2021akh}.

These reflections suggest that BHs could be surrounded by a thin atmosphere of width $\epsilon$, formed by a $\delta$-source at the Schwarzschild horizon and the strong gravitational field near the horizon. We can attempt to estimate the thickness of this atmosphere as the distance from the Schwarzschild radius to the closest photon sphere, obtained from the geodesic equations in the strong-field limit.

The classical trajectories of particles can be derived from the wavefunctions of quantum particles in the eikonal approximation \cite{Gold}. The Klein-Gordon wavefunctions in \eqref{wave}, which are related to classical motion, formally satisfy the relativistic Hamilton-Jacobi equation \cite{Motz:1964dvt}:
\begin{equation} \label{geodesic}
g_{\mu\nu}p^\mu p^\nu - \mu^2 = 0~,
\end{equation}
where $p^\nu = \mu \, dx^\nu/ds$ represents the relativistic 4-momentum. In the quasi-classical approximation, the scalar wave function in \eqref{wave} can be written in terms of an amplitude and phase:
\begin{equation} \label{Phi=rho}
\Phi = A e^{iS}~,
\end{equation}
where $S(x^\nu)$ is the Hamilton principal function, typically used to define classical momentum $p^\nu \sim \nabla^\nu S$. In the eikonal approximation, both $S(x^\nu)$ and $A(x^\nu)$ are treated as real. Inserting \eqref{Phi=rho} into \eqref{wave} gives:
\begin{equation} \label{wave-2}
- A \nabla_\nu S \nabla^\nu S + \Box A + 2i \nabla_\nu A \nabla^\nu S + i A \Box S + \mu^2 = 0~.
\end{equation}
With both $A$ and $S$ being real, we obtain the system of equations:
\begin{equation} \label{KG-system}
\begin{split}
A \Box S + 2 \nabla_\nu S \nabla^\nu A = 0~, \\
\Box A - A \nabla_\nu S \nabla^\nu S + \mu^2 A = 0~.
\end{split}
\end{equation}
In the eikonal approximation, the wave amplitude $A$ is typically assumed to vary slowly, while the Hamilton principal function $S$ is large enough for the most significant term in \eqref{wave-2} to be the one quadratic in $S$.

As shown previously, in the $\epsilon$-width shell around the BH horizon, the amplitude behaves as an exponentially time-dependent function \eqref{T}:
\begin{equation} \label{A}
A \sim e^{\pm t/r_s}~.
\end{equation}
The negative sign in the amplitude corresponds to the decay mode, or absorption of particles by the BH. Of greater interest is the case with the increasing amplitude in \eqref{A}, corresponding to the scattering of particles by the BH horizon. These reflecting particles contribute to forming the BH atmosphere with a width $\epsilon$. To estimate $\epsilon$, we substitute the increasing solution from \eqref{A} into the system \eqref{KG-system}:
\begin{equation} \label{KG-system-2}
\begin{split}
\Box S + \frac {2}{r_s} g^{tt} \dot S = 0~, \\
\frac {1}{r_s^2} g^{tt} - \nabla_\nu S \nabla^\nu S + \mu^2 = 0~,
\end{split}
\end{equation}
where the overdot denotes a time derivative. Solving the first equation of this system gives:
\begin{equation} \label{S}
S = E \int^t e^{-2t/r_s}dt + \int^x p^i dx_i ~,
\end{equation}
where $E$ and $p^i$ are constants, and $i$ indexes the three space-like coordinates.

The exponential warp factor in \eqref{S} suggests a reduction in the speed of light near the BH horizon, which leads to quasi-stable orbits in the thin atmosphere. However, these orbits are unstable, meaning a particle cannot remain in orbit indefinitely, even under ideal conditions. Estimates for the number of loops on the photon sphere suggest $n \leq 3$ \cite{Sneppen:2021taq, Chandra}. Assuming $n = 2$ for the average number of loops of particles in the BH atmosphere, each lasting approximately $t \sim 2\pi r_S$, and neglecting the time dependence of the particle's energy in \eqref{S}, we obtain:
\begin{equation}
p^0 = e^{-4\pi n} E \approx e^{-25}E \sim 10^{-11} E~,
\end{equation}
indicating an effective reduction in the speed of light by a factor of $\sim 10^{-11}$. We then identify:
\begin{equation}
\nabla^\alpha S = p^\alpha = \mu \frac {dx^\alpha}{ds}~,
\end{equation}
where $p^0 = 10^{-11}\,E$ and $x^0 = 10^{-11}\,t$.

Taking the $\nabla_\alpha$ derivative of the second equation in \eqref{KG-system-2} and applying the metric compatibility condition $\nabla_\alpha g^{\mu\nu} = 0$, we obtain:
\begin{equation} \label{}
\nabla_\nu S \nabla^\nu \nabla_\alpha S = 0 ~.
\end{equation}

Using the relation:
\begin{equation}
\frac {dx^\nu}{ds}\, \frac {\partial}{\partial x^\nu} = \frac {d}{ds}~,
\end{equation}
we recover the geodesic equation \cite{Mannheim:2021fql}:
\begin{equation}
\mu^2 \left( \frac {d^2x^\alpha}{ds^2} + \Gamma^\alpha_{\mu\nu} \frac {dx^\mu}{ds}\frac {dx^\nu}{ds}\right) = 0~,
\end{equation}
from the Klein-Gordon equation \eqref{wave}. For geodesics with constant $r$ and $\theta$, we obtain:
\begin{equation}
\frac{dr}{ds} = \frac{d^{2}r}{ds^{2}} = \frac{d\theta}{ds} = 0~,
\end{equation}
and from the radial geodesic equation, we deduce:
\begin{equation} \label{dphi/dt-1}
10^{-12}dt^2 = \frac{2r^3}{r_s}\sin^2\theta d\phi^2~.
\end{equation}

To derive the radius of quasi-classical orbits near the BH horizon, as for the classical photon sphere, we set $ds = dr = d\theta = 0$ in the Schwarzschild metric:
\begin{equation} \label{dphi/dt-2}
\left(1 - \frac{r_s}{r}\right)dt^2 = r^2 \sin^2\theta d\phi^2~.
\end{equation}
The relation $d\phi/dt$ is then derived using the radial geodesic equation \eqref{dphi/dt-1} with the effectively reduced speed of light. From \eqref{dphi/dt-1} and \eqref{dphi/dt-2}, we obtain:
\begin{equation}
10^{-22} \, \frac{r_s}{2r} = 1 - \frac{r_s}{r}~.
\end{equation}
Thus, the width of the BH's "classical atmosphere" is estimated as:
\begin{equation}
\epsilon = r - r_s \sim 10^{-22}\, r_s ~,
\end{equation}
which for supermassive BHs is consistent with results from entropy calculations in the brick-wall model \cite{Ghosh:2009xg}. The emergence of the black hole atmosphere suggests that the region near the horizon is highly dynamic, with particles undergoing rapid absorption and reflection.


\section{Conclusions} \label{Concl}

In this paper, we investigated the behavior of matter fields near the Schwarzschild horizon, focusing on solutions to the Klein-Gordon equation in the quasi-classical approximation. Our analysis shows that quantum particles near the horizon exhibit exponentially decaying (or growing) time-dependent solutions with a complex phase, rather than the periodic-in-time solutions typically associated with particles in black hole exteriors or interiors. This suggests that particles are unlikely to cross the Schwarzschild horizon; instead, they are either absorbed by the black hole or reflected back, contributing to the formation of a thin atmosphere around the horizon.

Using the Klein-Gordon equation, we derived the geodesic equations in the quasi-classical limit and estimated the width of this black hole atmosphere. Our results indicate that the atmosphere has a width of approximately $\epsilon \sim 10^{-22} r_s$, where $r_s$ is the Schwarzschild radius. This estimate aligns with previous findings based on entropy calculations in the brick-wall model, further supporting the idea that black holes are surrounded by a dynamic, quantum-mechanical layer of reflected particles.

The implications of our findings extend beyond the classical description of black holes. The existence of a black hole atmosphere suggests that the near-horizon region is a site of intense quantum activity, where particles undergo rapid absorption and reflection processes. This could have significant consequences for our understanding of black hole thermodynamics, quantum gravity, and the information paradox. For instance, one could explore the possibility that massless fermions trapped in the (1+2)-dimensional layer around a black hole horizon might acquire mass, thereby breaking parity and time-reversal symmetries \cite{Dunne:1998qy, Wilczek}. In (1+2) dimensions, a fermion can also perturbatively generate a Chern-Simons term for a gauge field, further breaking these symmetries \cite{Deser:1981wh, Redlich:1983dv}. Since $CPT$ symmetry holds, this could lead to effective $C$ and $CP$ violation, which might enhance baryon number violation processes near the horizon, as the other Sakharov conditions are also strengthened by the intense gravitational field of the black hole atmosphere \cite{Gogberashvili:2024ogp}.

Future work could investigate observational signatures of the black hole atmosphere, particularly in gravitational wave emissions and Hawking radiation. Additionally, exploring higher-order quantum corrections and the interplay between quantum fields and strong gravitational fields near the horizon may offer deeper insights into the quantum nature of black holes and their role in the broader framework of quantum gravity.



\begin{thebibliography}{99}

\bibitem{BH-1} S.~Chandrasekhar,
              {\it The Mathematical Theory of Black Holes}
              (Clarendon, New York 1983).

\bibitem{BH-2} S.~Carroll,
              {\it Spacetime and Geometry: An Introduction to General Relativity}
              (Addison-Wesley, San Francisco 2004).

\bibitem{BH-3} E.~Poisson,
              {\it A Relativist's Toolkit: The Mathematics of Black-Hole Mechanics}
              (Cambridge University Press, Cambridge 2004).

\bibitem{Wald} R.~M.~Wald,
              {\it General Relativity}
              (University of Chicago Press: Chicago 1984.

\bibitem{gravMTW} C.~W.~Misner, K.~S.~Thorne and J.~A.~Wheeler,
                 {\it Gravitation}
                 (Freeman, San Francisco 1973).

\bibitem{Ansoldi:2008jw} S.~Ansoldi,
``Spherical black holes with regular center: A Review of existing models including a recent realization with Gaussian sources,''
[arXiv: 0802.0330 [gr-qc]].

\bibitem{Cardoso:2019rvt} V.~Cardoso and P.~Pani,
``Testing the nature of dark compact objects: a status report,''
Living Rev. Rel. \textbf{22} (2019) 4,
doi: 10.1007/s41114-019-0020-4
[arXiv: 1904.05363 [gr-qc]].

\bibitem{Malafarina:2017csn} D.~Malafarina,
``Classical collapse to black holes and quantum bounces: A review,''
Universe \textbf{3} (2017) 48,
doi: 10.3390/universe3020048
[arXiv: 1703.04138 [gr-qc]].

\bibitem{Kerr:2023rpn} R.~P.~Kerr,
``Do black holes have singularities?,''
[arXiv: 2312.00841 [gr-qc]].

\bibitem{Modif-1} S.~B.~Giddings,
``Black hole information, unitarity, and nonlocality,''
Phys. Rev. D \textbf{74} (2006) 106005,
doi: 10.1103/PhysRevD.74.106005
[arXiv: hep-th/0605196 [hep-th]].

\bibitem{Modif-3} J.~Maldacena and L.~Susskind,
``Cool horizons for entangled black holes,''
Fortsch. Phys. \textbf{61} (2013) 781,
doi: 10.1002/prop.201300020
[arXiv: 1306.0533 [hep-th]].

\bibitem{Howard:1984qp} K.~W.~Howard and P.~Candelas,
``Quantum stress tensor in Schwarzschild space-time,''
Phys. Rev. Lett. \textbf{53} (1984) 403,
doi: 10.1103/PhysRevLett.53.403.

\bibitem{tHooft:1984kcu} G.~'t Hooft,
``On the quantum structure of a black hole,''
Nucl. Phys. B \textbf{256} (1985) 727,
doi: 10.1016/0550-3213(85)90418-3.

\bibitem{Mathur:2005zp} S.~D.~Mathur,
``The Fuzzball proposal for black holes: An Elementary review,''
Fortsch. Phys. \textbf{53} (2005) 793,
doi: 10.1002/prop.200410203
[arXiv: hep-th/0502050 [hep-th]].

\bibitem{Modif-2} A.~Almheiri, D.~Marolf, J.~Polchinski and J.~Sully,
``Black holes: Complementarity or firewalls?,''
JHEP \textbf{02} (2013) 062,
doi: 10.1007/JHEP02(2013)062
[arXiv: 1207.3123 [hep-th]].

\bibitem{meG1} M.~Gogberashvili and L.~Pantskhava,
``Black hole information problem and wave bursts,''
Int. J. Theor. Phys. \textbf{57} (2018) 1763,
doi: 10.1007/s10773-018-3702-x
[arXiv: 1608.04595 [physics.gen-ph]].

\bibitem{meG2} M.~Gogberashvili,
``Can quantum particles cross a horizon?,''
Int. J. Theor. Phys. \textbf{58} (2019) 3711,
doi: 10.1007/s10773-019-04242-0
[arXiv: 1712.02637 [gr-qc]].

\bibitem{meG3} R.~Beradze, M.~Gogberashvili and L.~Pantskhava,
``Reflective black holes,''
Mod. Phys. Lett. A \textbf{36} (2021) 2150200,
doi: 10.1142/S021773232150200X.

\bibitem{Kuchiev:2003fy} M.~Y.~Kuchiev and V.~V.~Flambaum,
``Scattering of scalar particles by a black hole,''
Phys. Rev. D \textbf{70} (2004) 044022,
doi: 10.1103/PhysRevD.70.044022
[arXiv: gr-qc/0312065].

\bibitem{Ghosh:2009xg} K.~Ghosh,
``Near-horizon geometry and the entropy of a minimally coupled scalar field in the Schwarzschild black hole,''
J. Phys. Soc. Jap. \textbf{85} (2016) 014101,
doi: 10.7566/JPSJ.85.014101
[arXiv: 0902.1601 [gr-qc]].


\bibitem{Colombeau-1} J.~F.~Colombeau,
                      {\it New Generalized Functions and Multiplication of Distributions}
                      (North Holland, Amsterdam 1984).

\bibitem{Colombeau-2} J.~F.~Colombeau,
                      {\it Elementary Introduction to New Generalized Functions}
                      (North Holland, Amsterdam 1985).

\bibitem{Gel-Sch} I.~M.~Gelfand and G.~E.~Schilov,
                  {\it Generalized Functions. Vol. I: Properties and Operations}
                  (Academic Press, New York - London 1964).

\bibitem{Heinzle:2001bk} J.~M.~Heinzle and R.~Steinbauer,
``Remarks on the distributional Schwarzschild geometry,''
J. Math. Phys. \textbf{43} (2002) 1493,
doi: 10.1063/1.1448684
[arXiv: gr-qc/0112047 [gr-qc]].

\bibitem{Foukzon} J.~Foukzon, E.~R.~Men'kova, A.~A.~Potapov and S.~A.~Podosenov,
``Was Polchinski wrong? Colombeau distributional Rindler space-time with distributional Levi-Civit\`a connection induced vacuum dominance. Unruh effect revisited,''
J. Phys. Conf. Ser. \textbf{1141} (2018) 012100,
doi: 10.1088/1742-6596/1141/1/012100.

\bibitem{Gogberashvili:2024ogp} M.~Gogberashvili and A.~S.~Sakharov,
``Black holes and baryon number violation: Unveiling the origins of early galaxies and the low-mass gap,''
Galaxies \textbf{13} (2025) 4,
doi: 10.3390/galaxies13010004
[arXiv: 2411.10847 [hep-ph]].

\bibitem{Ha} Y.~K.~Ha,
``External energy paradigm for black holes,''
Int. J. Mod. Phys. A \textbf{33} (2018) 1844025,
doi: 10.1142/S0217751X18440256
arXiv: 1811.02890 [physics.gen-ph].

\bibitem{Generalize} F.~Farassat,
                    {\it Introduction to Generalized Functions with Applications in Aerodynamics and Aeroacoustics}
                    (Langley Research Center, Virginia 1994).

\bibitem{meG4} M.~Gogberashvili,
``Einstein's hole argument and Schwarzschild singularities,''
Annals Phys. \textbf{452} (2023) 169274,
doi: 10.1016/j.aop.2023.169274
[arXiv: 2303.10348 [gr-qc]].


\bibitem{Matz} R.~A.~Matzner,
``Scattering of massless scalar waves by a Schwarzschild 'singularity',''
J. Mat. Phys. \textbf{9} (1968) 163,
doi: 10.1063/1.1664470.

\bibitem{Star} A.~A.~Starobinskii,
``Amplification of waves during reflection from a rotating 'black hole',''
Sov. Phys. JETP \textbf{37} (1973) 28.

\bibitem{Gogberashvili:2016xcx} M.~Gogberashvili and L.~Pantskhava,
``Black hole information problem and wave bursts,''
Int. J. Theor. Phys. \textbf{57} (2018) 1763,
doi: 10.1007/s10773-018-3702-x
[arXiv: 1608.04595 [physics.gen-ph]].

\bibitem{Qin} Y.-P.~Qin,
``Exact solutions to the Klein-Gordon equation in the vicinity of Schwarzschild black holes,''
Sci. China: Phys. Mech. Astron. {\bf 55} (2012) 381.


\bibitem{BH-wave-1} T.~Damour and R.~Ruffini,
``Black-hole evaporation in the Klein-Sauter-Heisenberg-Euler formalism,''
Phys. Rev. D \textbf{14} (1976) 332,
doi: 10.1103/PhysRevD.14.332.

\bibitem{BH-wave-2} S.~Sannan,
``Heuristic derivation of the probability distributions of particles emitted by a black hole,''
Gen. Rel. Grav. \textbf{20} (1988) 239,
doi: 10.1007/BF00759183.

\bibitem{BH-out} E.~Elizalde,
``Series solutions for the Klein-Gordon equation in Schwarzschild space-time,''
Phys. Rev. D \textbf{36} (1987) 1269,
doi: 10.1103/PhysRevD.36.1269.

\bibitem{Tunneling-1} K.~Srinivasan and T.~Padmanabhan,
``Particle production and complex path analysis,''
Phys. Rev. D \textbf{60} (1999) 024007.
doi: 10.1103/PhysRevD.60.024007
[arXiv: gr-qc/9812028].

\bibitem{Tunneling-2} E.~T.~Akhmedov, V.~Akhmedova and D.~Singleton,
``Hawking temperature in the tunneling picture,''
Phys. Lett. B \textbf{642} (2006) 124,
doi: 10.1016/j.physletb.2006.09.028
[arXiv: hep-th/0608098].

\bibitem{Beradze:2021akh} R.~Beradze, M.~Gogberashvili and L.~Pantskhava,
``Reflective black holes,''
Mod. Phys. Lett. A \textbf{36} (2021) 2150200,
doi: 10.1142/S021773232150200X.

\bibitem{Gogberashvili:2017xti} M.~Gogberashvili,
``Can Quantum Particles Cross a Horizon?,''
Int. J. Theor. Phys. \textbf{58} (2019) 3711,
doi: 10.1007/s10773-019-04242-0
[arXiv: 1712.02637 [gr-qc]].

\bibitem{Gold} H.~Goldstein,
              {\it Classical Mechanics}
              (Addison-Wesley, New York 1950).

\bibitem{Motz:1964dvt} L.~Motz and A.~Selzer,
``Quantum mechanics and the relativistic Hamilton-Jacobi equation,''
Phys. Rev. \textbf{133} (1964) B1622,
doi: 10.1103/physrev.133.b1622.

\bibitem{Sneppen:2021taq} A.~Sneppen,
``Divergent reflections around the photon sphere of a black hole,''
Sci. Rep. \textbf{11} (2021) 14247,
[erratum: Sci. Rep. \textbf{11} (2021) 17654]
doi: 10.1038/s41598-021-97272-w
[arXiv: 2107.04044 [gr-qc]].

\bibitem{Chandra} S.~Chandrasekhar,
                  {\it The Mathematical Theory of Black Holes}
                  (Oxford university press, Oxford 1998).

\bibitem{Mannheim:2021fql} P.~D.~Mannheim,
``Critique of the use of geodesics in astrophysics and cosmology,''
Class. Quant. Grav. \textbf{39} (2022) 245001,
doi: 10.1088/1361-6382/ac8140
[arXiv: 2105.08556 [gr-qc]].


\bibitem{Dunne:1998qy}  G.~V.~Dunne,
``Aspects of Chern-Simons theory,''
[arXiv: hep-th/9902115 [hep-th]].

\bibitem{Wilczek} F.~Wilczek,
                  {\it Fractional Statistics and Anyon Superconductivity}
                  (World Scientific, Singapore 1990).

\bibitem{Deser:1981wh} S.~Deser, R.~Jackiw and S.~Templeton,
``Topologically massive gauge theories,''
Annals Phys. \textbf{281} (2000) 409,
doi: 10.1006/aphy.2000.6013.

\bibitem{Redlich:1983dv} A.~N.~Redlich,
``Parity violation and gauge noninvariance of the effective gauge field action in three-dimensions,''
Phys. Rev. D \textbf{29} (1984) 2366,
doi: 10.1103/PhysRevD.29.2366.

\end{thebibliography}
\end{document}